\documentclass[twoside,twocolumn,9pt]{article}
\usepackage{extsizes}
\usepackage[super,sort&compress,comma]{natbib} 
\usepackage[version=3]{mhchem}
\usepackage[left=1.5cm, right=1.5cm, top=1.785cm, bottom=2.0cm]{geometry}
\usepackage{balance}
\usepackage{mathptmx}
\usepackage{sectsty}
\usepackage{graphicx} 
\usepackage{lastpage}
\usepackage[format=plain,justification=justified,singlelinecheck=false,font={stretch=1.125,small,sf},labelfont=bf,labelsep=space]{caption}
\usepackage{float}
\usepackage{fancyhdr}
\usepackage{fnpos}
\usepackage[english]{babel}
\addto{\captionsenglish}{%
  
}
\usepackage{array}
\usepackage{droidsans}
\usepackage{charter}
\usepackage[T1]{fontenc}
\usepackage[usenames,dvipsnames]{xcolor}
\usepackage{setspace}
\usepackage[compact]{titlesec}
\usepackage{hyperref}
\usepackage{amsmath}

\usepackage{ulem}

\usepackage{epstopdf}

\definecolor{cream}{RGB}{222,217,201}

\begin{document}
\pagestyle{fancy}
\thispagestyle{plain}
\fancypagestyle{plain}{
\renewcommand{\headrulewidth}{0pt}
}

\makeFNbottom
\makeatletter
\renewcommand\LARGE{\@setfontsize\LARGE{15pt}{17}}
\renewcommand\Large{\@setfontsize\Large{12pt}{14}}
\renewcommand\large{\@setfontsize\large{10pt}{12}}
\renewcommand\footnotesize{\@setfontsize\footnotesize{7pt}{10}}
\makeatother

\renewcommand{\thefootnote}{\fnsymbol{footnote}}
\renewcommand\footnoterule{\vspace*{1pt}%
\color{cream}\hrule width 3.5in height 0.4pt \color{black}\vspace*{5pt}} 
\setcounter{secnumdepth}{5}

\makeatletter 
\renewcommand\@biblabel[1]{#1}            
\renewcommand\@makefntext[1]%
{\noindent\makebox[0pt][r]{\@thefnmark\,}#1}
\makeatother 
\renewcommand{\figurename}{\small{Fig.}~}
\sectionfont{\sffamily\Large}
\subsectionfont{\normalsize}
\subsubsectionfont{\bf}
\setstretch{1.125} 
\setlength{\skip\footins}{0.8cm}
\setlength{\footnotesep}{0.25cm}
\setlength{\jot}{10pt}
\titlespacing*{\section}{0pt}{4pt}{4pt}
\titlespacing*{\subsection}{0pt}{15pt}{1pt}

\fancyfoot{}
\fancyfoot[LO,RE]{\vspace{-7.1pt}\includegraphics[height=9pt]{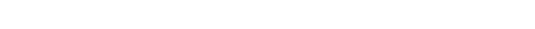}}
\fancyfoot[CO]{\vspace{-7.1pt}\hspace{13.2cm}\includegraphics{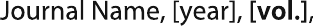}}
\fancyfoot[CE]{\vspace{-7.2pt}\hspace{-14.2cm}\includegraphics{head_foot/RF}}
\fancyfoot[RO]{\footnotesize{\sffamily{1--\pageref{LastPage} ~\textbar  \hspace{2pt}\thepage}}}
\fancyfoot[LE]{\footnotesize{\sffamily{\thepage~\textbar\hspace{3.45cm} 1--\pageref{LastPage}}}}
\fancyhead{}
\renewcommand{\headrulewidth}{0pt} 
\renewcommand{\footrulewidth}{0pt}
\setlength{\arrayrulewidth}{1pt}
\setlength{\columnsep}{6.5mm}
\setlength\bibsep{1pt}

\makeatletter 
\newlength{\figrulesep} 
\setlength{\figrulesep}{0.5\textfloatsep} 

\newcommand{\topfigrule}{\vspace*{-1pt}%
\noindent{\color{cream}\rule[-\figrulesep]{\columnwidth}{1.5pt}} }

\newcommand{\botfigrule}{\vspace*{-2pt}%
\noindent{\color{cream}\rule[\figrulesep]{\columnwidth}{1.5pt}} }

\newcommand{\dblfigrule}{\vspace*{-1pt}%
\noindent{\color{cream}\rule[-\figrulesep]{\textwidth}{1.5pt}} }

\makeatother

\twocolumn[
  \begin{@twocolumnfalse}
{\includegraphics[height=30pt]{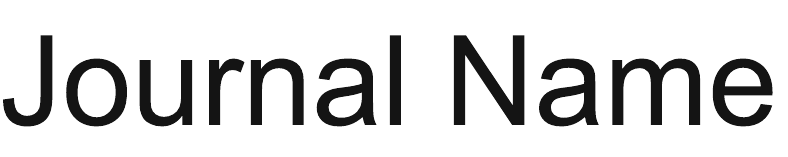}\hfill\raisebox{0pt}[0pt][0pt]{\includegraphics[height=55pt]{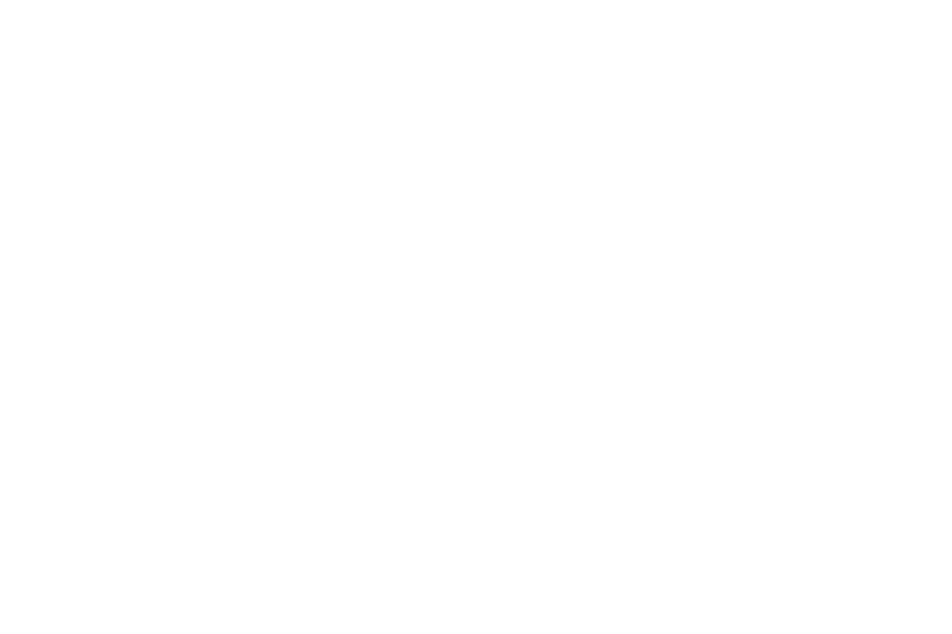}}\\[1ex]
\includegraphics[width=18.5cm]{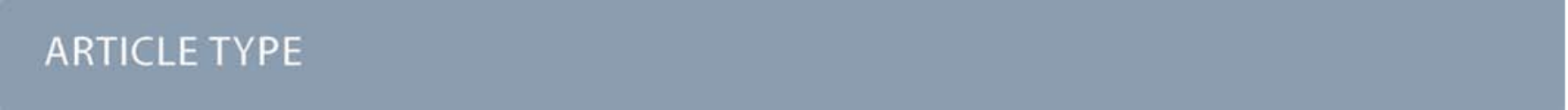}}\par
\vspace{1em}
\sffamily
\begin{tabular}{m{4.5cm} p{13.5cm} }

\includegraphics{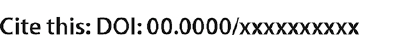} &
\noindent\LARGE{\textbf{Speeding up biphasic reactions with surface nanodroplets $^\dag$}} \\

\vspace{0.3cm} & \vspace{0.3cm} \\

 & \noindent\large{Zhengxin Li,\textit{$^{a}$} 
 Akihito Kiyama,\textit{$^b$} 
 Hongbo Zeng,$^{\ast}$\textit{$^{a}$} 
 Detlef Lohse,$^{\ast}$\textit{$^{c}$}
 and Xuehua Zhang$^{\ast}$\textit{$^{a,b,c}$}} \\

\includegraphics{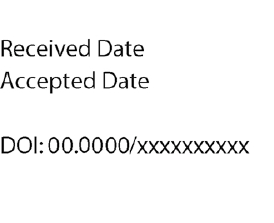} & \noindent\normalsize{Biphasic chemical reactions compartmentalized in small droplets offer advantages, such as streamlined procedures for chemical analysis, enhanced chemical reaction efficiency and high specificity of conversion.  In this work, we experimentally and theoretically investigate the rate for biphasic chemical reactions between acidic nanodroplets on a substrate surface and basic reactants in a surrounding bulk flow. The reaction rate is measured by droplet shrinkage as the product is removed from the droplets by the flow. In our experiments, we determine the dependence of the reaction rate on the flow rate and the solution concentration. 
The theoretical analysis predicts that the life time $\tau$ of the droplets scales with Peclet number $Pe$ and the reactant concentration in the bulk flow $c_{re,bulk}$ as $\tau\propto Pe^{-3/2}c_{re,bulk}^{-1}$, in good agreement with our experimental results. Furthermore, we found that the product from the reaction on an upstream surface can postpone the droplet reaction on a downstream surface, possibly due to the adsorption of interface-active products on the droplets in the downstream. The time of the delay decreases with increasing $Pe$ of the flow and also with increasing reactant concentration in the flow, following the  scaling same as that of the reaction rate with these two parameters.  Our findings provide insight for  the ultimate aim to enhance droplet reactions under flow conditions.} \\

\end{tabular}

 \end{@twocolumnfalse} \vspace{0.6cm}

 ]

\renewcommand*\rmdefault{bch}\normalfont\upshape
\rmfamily
\section*{}
\vspace{-1cm}


\footnotetext{\textit{$^a$ Department of Chemical and Materials Engineering, Faculty of Engineering, University of Alberta, Edmonton, Alberta, T6G 1H9, Canada.}}
\footnotetext{\textit{$^b$ Institute of Global Innovation Research, Tokyo University of Agriculture and Technology, Nakacho 2-24-16 Koganei, Tokyo, 184-8588, Japan}}
\footnotetext{\textit{$^c$ Physics of Fluids Group, Max Planck Center Twente for Complex Fluid Dynamics, JM
Burgers Center for Fluid Dynamics, Mesa+, Department of Science and Technology,
University of Twente, Enschede 7522 NB, The Netherlands}}

\footnotetext{\dag~Electronic Supplementary Information (ESI) available: [details of any supplementary information available should be included here]. See DOI: 00.0000/00000000.}


\section{Introduction}
Small-size droplets are omnipresent in nature and technology, including lab-on-chip, emulsions, aerosols, sneezing and coughing, cell metabolism compartments, heterogeneous catalysis, polymer synthesis, micro-extraction, among many others.\cite{lohse2015surface,thiam2013biophysics,li2017enhancement,liu2002synthesis,lin2016magnetic,okubo1980estimation,wang2019autonomic,qian2019surface}
At present, ‘on-droplet’ chemistry attracts increasing research attention.\cite{bain2017droplet} 
Chemical reactions compartmentalized in small-size droplets can potentially be highly efficient with large throughputs due to their high surface area-to-volume ratio and their discrete nature.\cite{kelly2007miniaturizing,wang2009efficient,feng2018droplet,guardingo2016reactions}
Noticeably, the kinetics of chemical reactions on the droplet surface can be significantly enhanced.\cite{yan2016organic} 
For instance, chemical reactions in aerosol droplets are  accelerated, in some cases even by a factor of $10^6$ compared to their bulk counterparts.\cite{lee2015acceleration,lee2015microdroplet} 
Acceleration can also be found in a diverse range of biphasic reactions in confinement that involves two immiscible fluids, such as in micro-sized emulsion droplets, thin liquid films, inverted micelles or at the surface of aerosol particles.\cite{wei2017reaction,griffith2012situ,wiebenga2016nanoconfinement} 
Apart from the reaction kinetics, the intermediates or products from droplet reactions can also be different from those from the counterparts in bulk.\cite{lee2018spontaneous} Some reactions that are impossible without catalysts in the bulk can take place spontaneously in droplets. For example, Nam et al. demonstrated that phosphorylation of sugars occurred spontaneously in aqueous microdroplets.\cite{nam2017abiotic}
The shifted balance and accelerated kinetics of reactions in small droplets may demonstrate a plausible route to the production of complex biomolecules outside of living systems.\cite{fallah2014enhanced} Reactions confined in small droplets have been proposed to explain how synthetic reaction for complex biomolecules that are thermodynamically unfavorable in aqueous bulk could occur in the origin of life on early earth.\cite{vaida2017prebiotic,urban2014compartmentalised}
 
\par The mechanisms for reaction acceleration in droplets are still unclear so far.
Two possible explanations are proposed in the literature. 
(1) In case of flying droplets in air created from electrospraying, the solvent in the droplet may evaporate, leading to rapid shrinkage in droplets size and the increase in the concentration of reagents inside droplets. 
In addition, reagent diffusion is quick in small size droplets, which may contribute to the enhanced reaction kinetics. \cite{carroll2013experimental,lee2015microdroplet} 
(2) Another important effect may be from the large surface area-to-volume ratio of droplets, compared to larger drops or the bulk liquid. 
Nakatani and co-workers found that electron transfer was accelerated at droplet surface.\cite{nakatani1995direct,nakatani1995droplet,nakatani1996electrochemical}  Furthermore, Fallah-Araghi et al.\cite{fallah2014enhanced} demonstrated that the reaction rate is inversely proportional to the droplet radius, related to the preference of product adsorption and desorption at the droplet interface that accelerates the rate and shifts the balance of the chemical reaction.\cite{fallah2014enhanced} 
The active energy barrier was found to be negligible for the reactions within droplets, possibly due to the molecular configuration of reactants at the droplet surface.\cite{nam2017abiotic}

How to distinguish between the relevance of these two suggested mechanisms? Immersed surface nanodroplets provide a unique platform for studying reaction kinetics under well-controlled conditions, eliminating the influence from concentrating effects due to solvent evaporation though one has dissolution effect.
These droplets have a maximal thickness from several to several hundred nanometers (namely nanodroplets) and a volume typically on the order of femto- or atto- liters, located on the solid surface in contact with a bulk liquid that is immiscible with the droplet liquid.\cite{lohse2015surface}
The size distribution and the number density of surface nanodroplets can be well controlled by solution composition and flow condition during a simple process of solvent exchange.\cite{zhang2015,lu2016influence}
The droplet morphology can be tailored by the properties and patterns of the substrates.\cite{peng2015spontaneous,bao2018}
The long term stability of surface nanodroplets due to their poor solubility enables us to track the reaction kinetics in-situ with  sufficient temporal and spatial resolution.

\par In this work, we investigate the rate of  biphasic reactions between surface nanodroplets and the reactant solution in an external flow. In our model systems, acidic droplets react with basic solution in the flow. 
The product from the reactions is surface active, carried away by the surrounding flow after desorption from the reacting droplets. The combined effects from the reaction and the mass loss of the product lead to the shrinkage of surface nanodroplets. The objective of this study is to improve the understanding of the chemical kinetics on the surfaces of small-sized droplets. The findings will be valuable to guide the design of droplet-based reactions in flows for heterogeneous catalysis, micro-extraction and other applications. 

\section{Methodology}
\subsection{Chemicals and materials}
Oleic acid (90\%, Fisher Scientific), ethanol (90\%, Fisher Scientific), octyldecyltrichlorosilane (OTS) (95\% Fisher Scientific) were used as received without further purification. Water was from Milli-Q  (18.2 M$\Omega$). 
Silicon substrates were hydrophobilized with OTS, prepared by following a procedure reported previously.\cite{lessel2015}
The OTS-coated substrates were cleaned by sonication in ethanol for 10 min and dried in a stream of air before use.

Oleic acid (OA) was chosen as the droplet liquid. 
To perform the solvent exchange, two solutions were prepared. 
The first solution (solution A) was 2.4\% (v/v) oleic acid in the mixture of ethanol and water, where the ratio was 6.5:3.5. 
The second solution (solution B) was water.
To trigger the chemical reaction of oleic droplets, solution C (sodium hydroxide aqueous solution) was also prepared.

\subsection{Formation of droplets}
\begin{figure}[htb!]
 \centering
 \includegraphics[height=11.5cm]{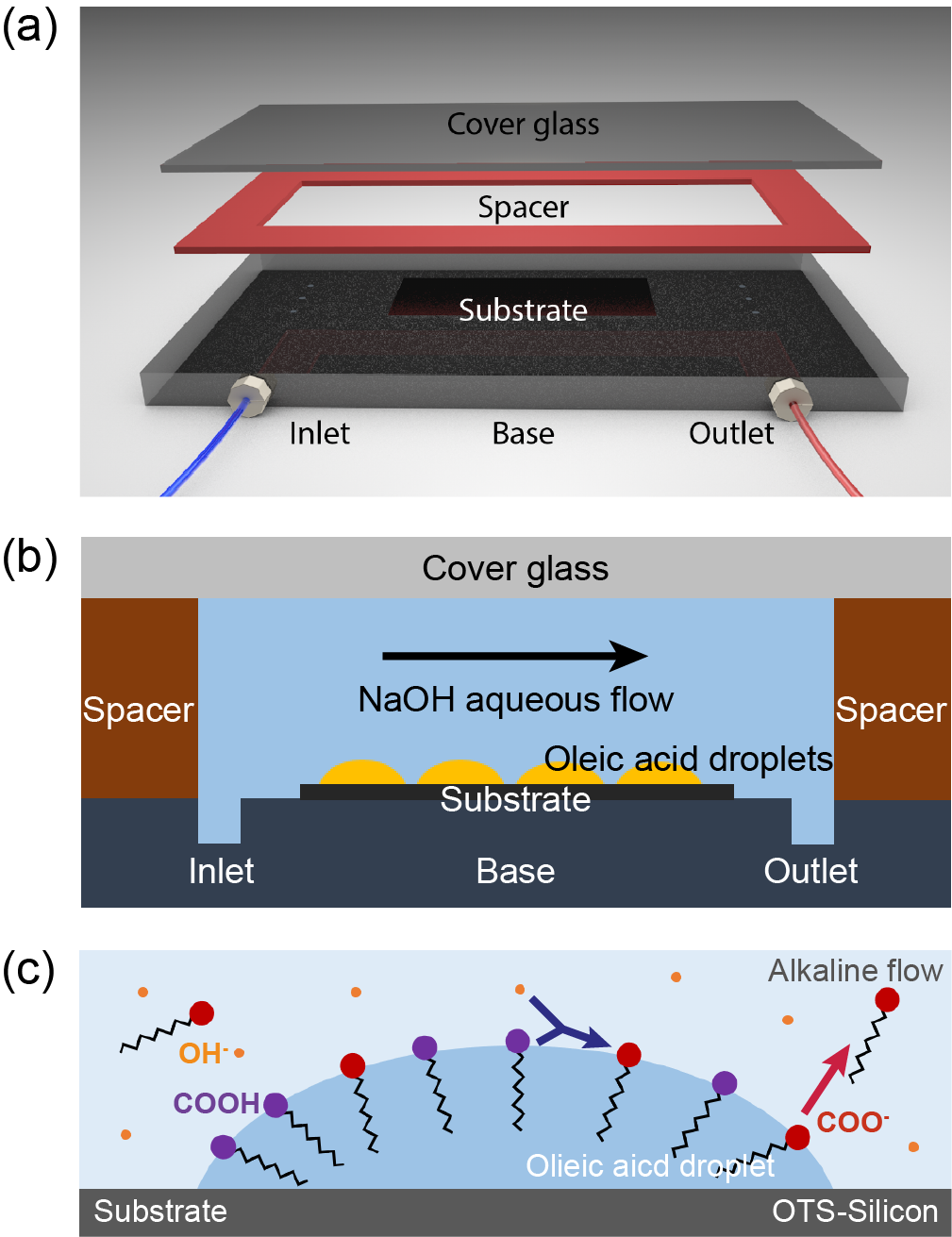}
  \caption{Schematics of the experimental set-up: (a) A fluid chamber for solvent exchange and droplets reaction. The chamber consists of a top cover glass, a spacer, and a base with inlet and outlet. The hydrophobic substrate was attached to the base. The height of the fluid chamber was adjusted by the thickness of the spacer. (b) The experimental set-up. Oleic acid (OA) droplets on the substrate were formed by solvent exchange, and the alkaline solution was introduced from the inlet. NaOH reacts with the COOH group of oleic acid at the droplet surface. (c) An oleic droplet with a base radius R reacting with the alkaline flow. The COOH group (purple head group in the sketch) reacts with OH$^{-}$(orange circle) in the alkaline flow, forming a COO$^{-}$ group (red dot). The reactant and the product have the same hydrophobic tail (black chain). The blue arrow indicates the reaction. The product oleate desorbs from the droplet surface, transported away by the flow as indicated by the red arrow.}
\label{fgr:Setup}
\end{figure}

The solvent exchange was used to prepare reactive oil droplets of oleic acid (see \S 2.2).
Solution A introduced inside a house-made fluid chamber was replaced by solution B.
The design and components of the fluid chamber are shown in Figure \ref{fgr:Setup}(a).
The height of the chamber (the distance between the substrate and the cover glass) was 0.4 mm, the width of the chamber was 15 mm and the length of the substrate was 25 mm in all experiments.
The solvent exchange was performed at 21$^\circ$C.
The injection of solution B was controlled at 500 $\mu$L/min in terms of volume flow rate, to keep droplets formed with consistent number density and size distribution.
The initial surface coverage, the ratio of the substrate's surface area taken up by OA droplets after the solvent exchange, was fixed about $\approx$ 4\% for all groups of experiments.

\subsection{Chemical reaction}
After formation of nanodroplets by the solvent exchange, solution C was introduced into the fluid chamber to initiate the reaction, as sketched in Fig. \ref{fgr:Setup}(b).
The schematic drawing of the reaction process in a single droplet is shown in Fig. \ref{fgr:Setup}(c).
Owning to the hydrophilic property, hydrophilic carboxyl groups of oleic acid at the interface tend to stay at the water side.
At the same time, hydroxide ions from the bulk are convected to the droplet interface by the flow, attack carboxyl groups in the water side and convert the acid to oleate.
Due to its high surface activity, oleate stays at the interface, and in a long term, gradually dissolves into the aqueous phase and removed by the flow. 

\subsection{Parameter space for experiments}
We conducted the experiments at different flow rate $Q$ controlled by a syringe pump.
{\it Q} is the volume flow rate of the alkaline flow.
The dimensionless Peclet number (\textit{Pe}) is defined as 

\begin{equation}
  Pe = \frac{\bar{U}h}{D} = \frac{Q}{wD}
\end{equation}
where $D$, $\bar U$, $h$ and $w$ are respectively the diffusion constant of oleic acid, the average linear flow rate, the channel height (400 $\mu$m) and the channel width (130 mm). To study the effect from the flow rate, the concentration of NaOH in the solution $c_{re,bulk}$ was fixed at $4.0\times10^{-4}$M while {\it Pe} varied from 5 to 198. The time required for the solution to reach the same region of the surface at different flow rate was predetermined in experiments by flowing the solution inside an empty fluid cell at the same flow rate.

To investigate the concentration effect, we kept the {\it Pe} number constant at 31, while the concentration of NaOH ($c_{re,bulk}$) in the solution varied from $1.0\times10^{-4}$ M to 0.1 M. 

\subsection{Characterization of droplet size}
Reaction processes were recorded {\it in-situ} by an upright microscope with video camera (Nikon, 10x objective lens, 0.24 $\mu$m/pixel, 15.0 fps).
White-light LED was applied to trace the surface of the substrate by bright field imaging. 
The filmed images were processed by ImageJ and analyzed frame-by-frame by self-written Matlab codes. 
Based on the binarized images, the surface coverage SC, the base radius R with time, and the characteristic lifetime $\tau$ are determined. 
Surface coverage of the droplets on the substrate was analyzed over an area of 0.34 mm$^2$ with around 2000 droplets.

\section{Results and discussion}
\subsection{Dependence of droplet reaction rate on $Pe$}
\begin{figure}[hbt!]
\centering
  \includegraphics[height=19.2cm]{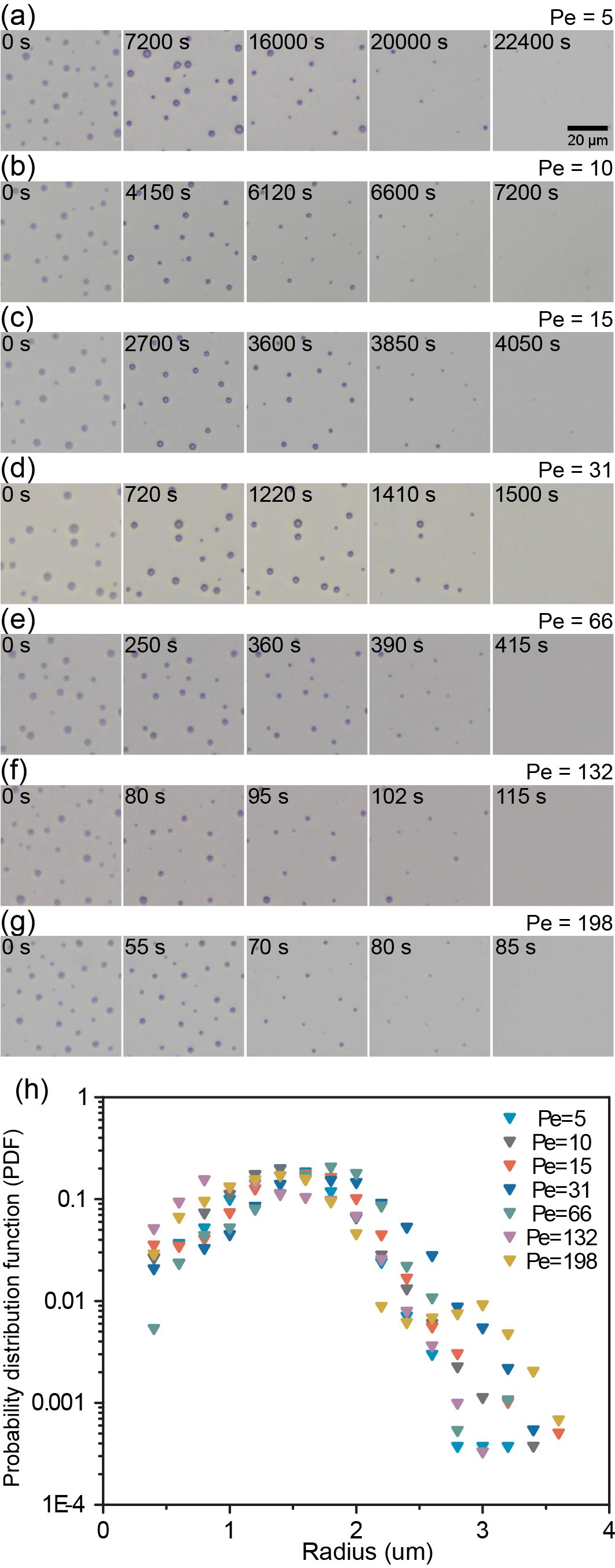}
  \caption{Optical images of reacting nanodroplets with alkaline solution of different flow rates. The frame rates of the videos are 15 fps, and the spatial resolution is 0.24 $\mu$m /pixel. The concentration of NaOH in each group was controlled at 4.0$\times$10$^{-4}$ M. The flow rates from (a) to (g) were respectively 17, 33, 50, 100, 215, 430 and 645 $\mu$L/min (corresponding Peclet numbers (Pe=Q/(wD)) thus were 5 to 198). The droplets in all images were produced by solvent exchange with exactly same flow and solution conditions to make initial droplets with consistent size and number density. (h) Probability distribution functions (PDF) of initial droplet sizes at different flow rates. Data shown are from droplets with the measured radius larger than 0.3 $\mu$m due to the spatial resolution.}
  \label{fgr:Pe1}
\end{figure}

In order to obtain consistent initial droplet conditions, all experiments were conducted from similar surface coverage ($\sim4\%$) of droplets on the substrate with averaged droplets radius of $\sim$ 1.2 $\mu$m.
Seven series of snapshots in Figure \ref{fgr:Pe1} reveals progressive shrinkage of reacting droplets from reacting with the alkaline flow at different flow rates. 
The time at the start of the reaction ($t_0$) shown in the first image of each row was defined as the moment when the flow of the solution enters the field of the view. 
Figure \ref{fgr:Pe1} (h) shows the probability distribution functions (PDF) at varying Peclet numbers. Results provided in plots suggest that initial droplet size distributions are highly consistent in the reactions with the basic solution supplied at different flow rates. Therefore any effect from different surface-to-volume ratios of the droplets on the reaction kinetic is expected to be same in all experiments.

\par Starting from $t_0$, droplets became darker in several seconds, which is seen by comparing images in the first and the second columns in Figure \ref{fgr:Pe1}. The darker color of the droplets was possibly due to the formation of the product that is surface active. Thus the shape of droplets may change and appear darker in images. In other words, the droplets were possibly covered by the product from the reaction.

\par After a period of time from the arrival of the alkaline solution, droplets started to shrink with a noticeable rate and eventually disappeared from the surface. 
The droplets dissolved faster as the Peclect number of the solution $Pe$ increases. In the two most apparent cases in Figure \ref{fgr:Pe1}(a)\&(g), at $Pe=198$ the droplets completely dissolved in 85 seconds $t$, but dissolution takes 6 hours at $Pe=5$. As comparison, droplets had not shrunk after 8 hrs under a continuous flow of pure water at 6 $ml/h$, showing that droplet shrinkage was not due to the dissolution of the droplet liquid, but due to the loss of the product from the chemical reaction. These results clearly demonstrate that the flow rate of the reactant solution has a significant impact on the rate of droplet reaction.

\begin{figure*}[htb!]
 \centering
 \includegraphics[height=10.2cm]{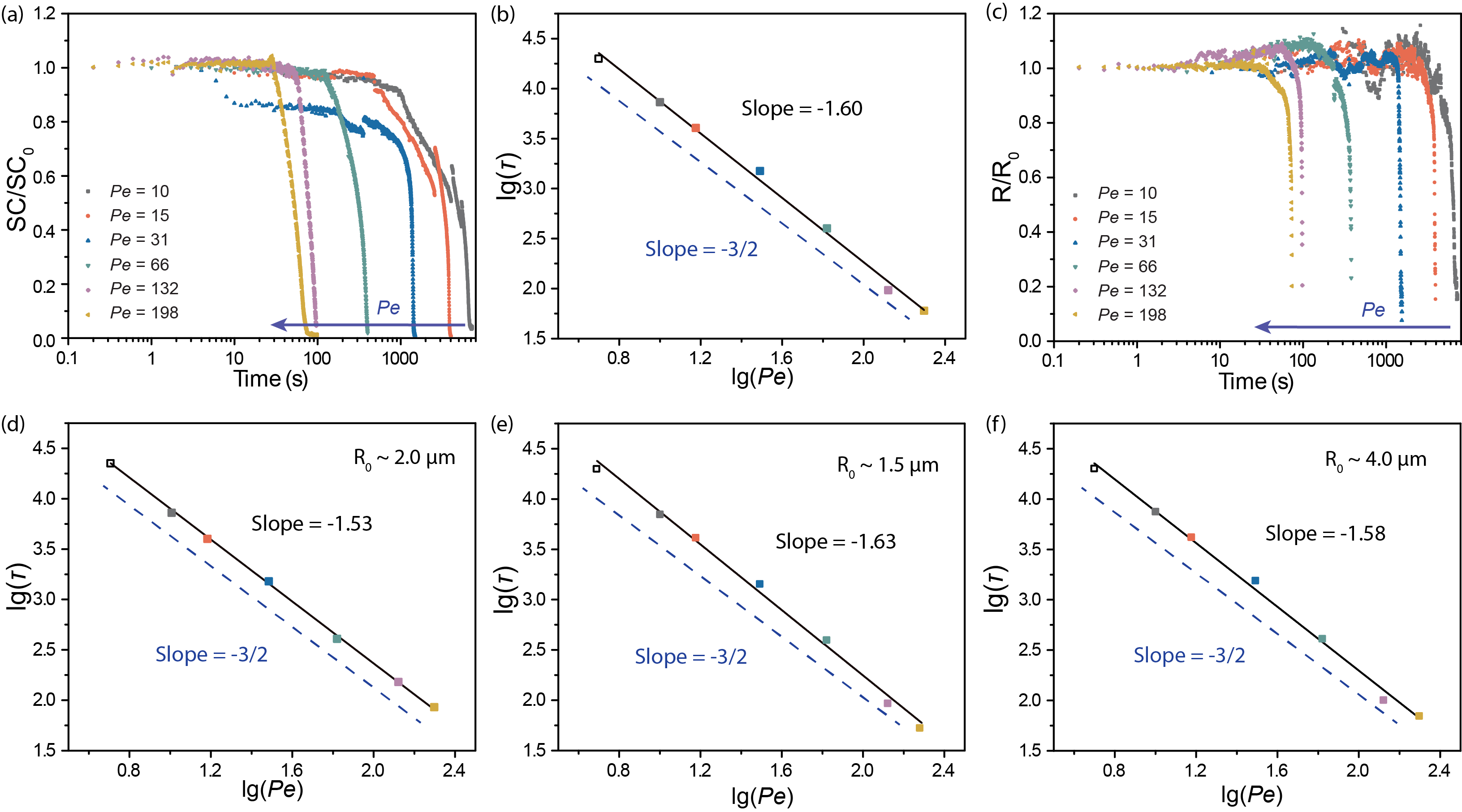}
 \caption{(a) Normalized surface coverage as function of time for droplet reaction at different flow rates. $SC$: surface coverage; $SC_0$: initial surface coverage. (b) Droplet lifetime $\tau$ as function of Peclet number. $\tau$ on the x-axis is the time required for $SC/SC_0$ to reach 0.1. (c) Normalized lateral radius as function of time for individual reacting droplets at different flow rates. Droplets presented here all have a similar initial size ($\sim$ 2.1 $\mu$m) and located at a similar location on the substrate. As the optical resolution of the microscope was 0.24 $\mu$m/pixel, the droplets once smaller than 0.24 $\mu$m in lateral diameters cannot be resolved. (d)$-$(f) Droplet lifetime $\tau$ based on the droplet radius as function of Peclet number. The initial radii of droplets in (d), (e), and (f) were around 2.1, 1.5, and 4.0 $\mu m$, respectively. The initial $\tau$ is the time required for $R/R_0$ to reach 0.1. The data in (a) and in (c) were obtained from the analysis of the images (b)$-$(g) in Fig.\ref{fgr:Pe1} and data in (b) and in (d)$-$(f) were from (a)$-$(g) in Fig.\ref{fgr:Pe1}. The black lines in (b) and (d)$-$(f) were obtained by fitting the experimental data, while the blue dashed lines represent the result $\tau\propto$ $Pe^{-3/2}$ from the scaling analysis.}
 \label{fgr:Pe2}
\end{figure*}

The quantitative analysis is shown in Fig. \ref{fgr:Pe2}(a) where the temporal surface coverage (normalized by surface coverage at $t_0$) is plotted as function of the Pe number of the alkaline flow.
A general feature is that there were two stages in the droplet dissolution after the arrival of the reacting solution in the flow. At an initial stage, the surface coverage and the droplet sizes did not experience significant shrinkage and showed somewhat flat responses with time. At the second stage, the droplets started to shrink with an increasing shrinkage rate.  The higher $Pe$ of the alkaline solution was, the earlier the surface coverage of the droplets started to decrease. 
The droplets continuously dissolved till they disappeared.
We note that a single exponential decay function cannot completely fit the curve of droplet dissolution. Especially at the late stage of dissolution, the slope of our experimental data is much faster than that of the fitting line generated by a single exponential decay function (Fig. S1). The reason is that droplet reactions take place in a background with product concentration varying with time, which will be analyzed and explained in detail later in \S 3.4.

Here we compare the difference in droplet lifetime for different $Pe$ numbers, measured by the shrinking rate in surface coverage  (Figure \ref{fgr:Pe2}(b)).
The colors of markers correspond to that in Figure \ref{fgr:Pe2}(a).

The lifetime of droplets $\tau$ here is defined as the time from the start of the reaction $t_0$ to the moment when surface coverage SC decreases to 10 \% of the initial surface coverage, SC$_0$.
Remarkably, all the data collapse into a single universal curve with a best fitting effective scaling exponent of $-1.60$, as indicated by the solid black line in the plot.
Note that the slope resulting from scaling analysis (slope=$-3/2$, presented later in \S 3.2) is also shown in the plot.
We also tried to analyze half of the original region (around 1000 droplets) and the analysis of fewer droplets can still yield consistent results (Fig. S2).

In addition to the surface coverage, the lateral sizes $R$ of individual droplets were also analyzed, namely its dependence on $Pe$. The initial radii $R_0$ of these droplets were all around 2.1 $\mu m$. We analyzed the lifetime of individual droplets $\tau$, as shown in Fig. \ref{fgr:Pe2}(c). Consistent with the two-stage reaction, as highlighted in Figure \ref{fgr:Pe2}(a), individual droplets also exhibit the feature of two-stage dissolution. 
The lifetime of individual droplets is shorter when the reacting solution is supplied at a faster flow rate. Here $\tau$ is the duration from the droplet in contact with the solution to the end of droplet dissolution. 
 
Fig. \ref{fgr:Pe2}(d) quantitatively shows the droplets lifetime $\tau$ of individual droplets as function of the $Pe$ number.
An effective scaling law of $\tau\propto$ $Pe^{-1.53}$ was found from the individual droplets, which fairly agrees with that found from the overall surface coverage (Figure \ref{fgr:Pe2}(b)).

To support that the results are unaffected by the droplet size, the lifetime of two additional groups of droplets were also analyzed at different $Pe$. The initial radii $R_0$ of droplets in these two groups were around 1.5 $\mu m$ and 4.0 $\mu m$ respectively. The effective scaling laws yield by these two groups were $Pe^{-1.63}$ and $Pe^{-1.58}$ (Fig. \ref{fgr:Pe2}(e)\&(f)), consistent with results in the 2.0 $\mu m$ initial radii group. Results from droplets with other initial sizes are shown in Fig. S3.

\subsection{Dependence of droplet reaction on NaOH concentration}
It is expected that the droplet dissolution rate is influenced by the concentration of the reacting solution. At identical flow conditions, we examined droplet reaction rates as the concentration of the alkaline solution was varied over three orders of magnitudes. 

\begin{figure}[htb!]
\centering
  \includegraphics[height=19.6cm]{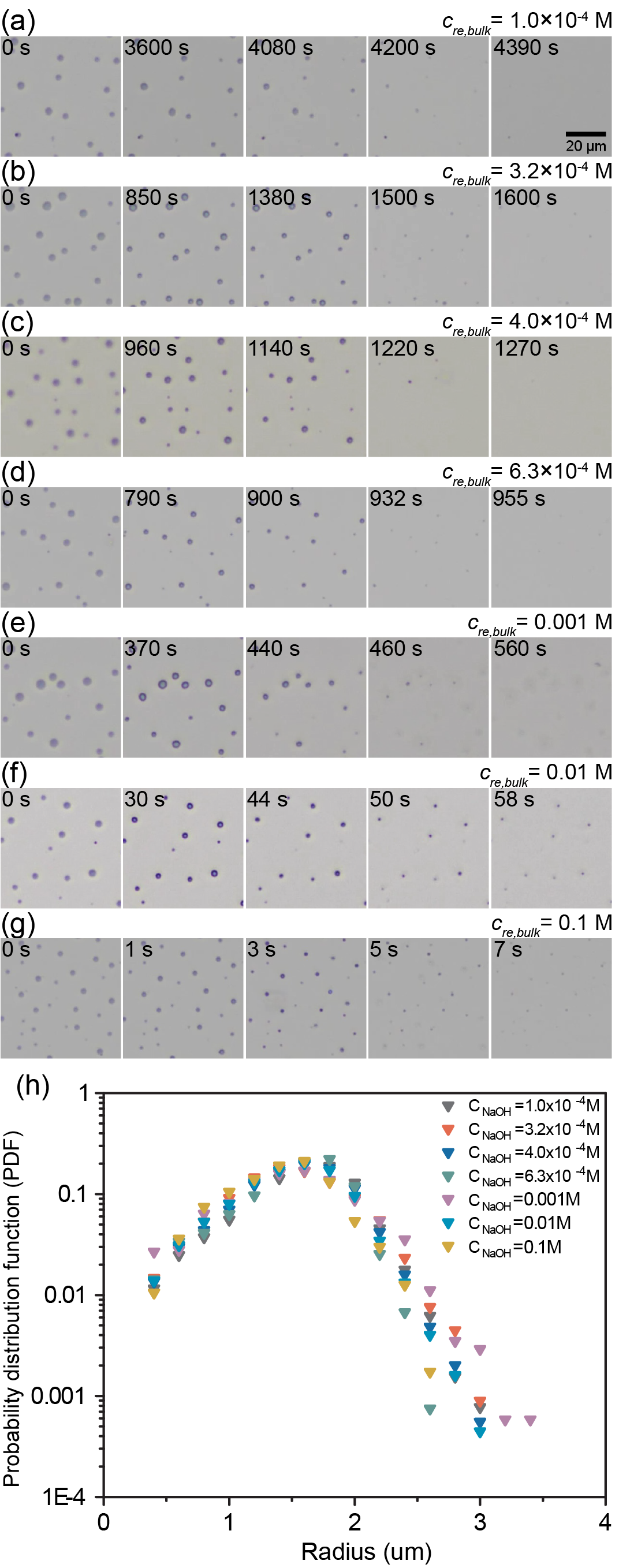}
  \caption{Optical images of surface nanodroplets reacting with the alkaline solution of different concentrations. The frame rate of videos was 15 fps, and the resolution was 0.24 $\mu$m /pixel. The flow rate for all groups from (a-g) was same, controlled at 100 $\mu$L/min ($Pe$=31). The concentration of alkaline in the solution $C_{re,bulk}$ was $1.0\times10^{-4}$M (a), $3.2\times10^{-4}$M (b), $4.0\times10^{-4}$M (c), $6.3\times10^{-4}$M (d), $0.001$M (e), $0.01$M (f), and $0.1$M (g). (h) Probability distribution functions (PDF) of initial droplet sizes at different NaOH concentrations. Data shown are from droplets with the measured radius larger than 0.3 $\mu$m.}
  \label{fgr:pH1}
\end{figure}

Figure \ref{fgr:pH1} displays microscopic images of oleic acid droplets dissolving in alkaline flow with different sodium hydroxide concentration. The difference in the lifetime of droplets shown in each row in Figure \ref{fgr:pH1} demonstrates that the droplets dissolve faster at a higher concentration of the alkaline solution.
The lifetime $\tau$ drops from $\sim$4400 seconds for $10^{-4}$M NaOH concentration (Figure \ref{fgr:pH1}(a)) to nearly 7 seconds for 0.1M NaOH concentration (Figure \ref{fgr:pH1}(g)).
Figure \ref{fgr:pH1} (h) demonstrates the probability distribution functions (PDF) at varying NaOH concentrations. Results provided in plots suggest that initial droplet size distributions are highly consistent in the reactions at different NaOH concentrations.

We noticed that for high NaOH concentrations, some residues were found at the end of the reaction (last column in Figure \ref{fgr:pH1}(e)$\sim$(g)). 
The residues with irregular shapes may be the product that did not dissolve into the alkaline flow immediately after the chemical reaction. 
At higher NaOH concentrations, the chemical reaction is very fast, and thus the local concentration of the product near the droplets becomes higher. The product cannot be transported by the flow immediately, resulting in residues left on the substrate.

\begin{figure*}[htb!]
\centering
  \includegraphics[height=7cm]{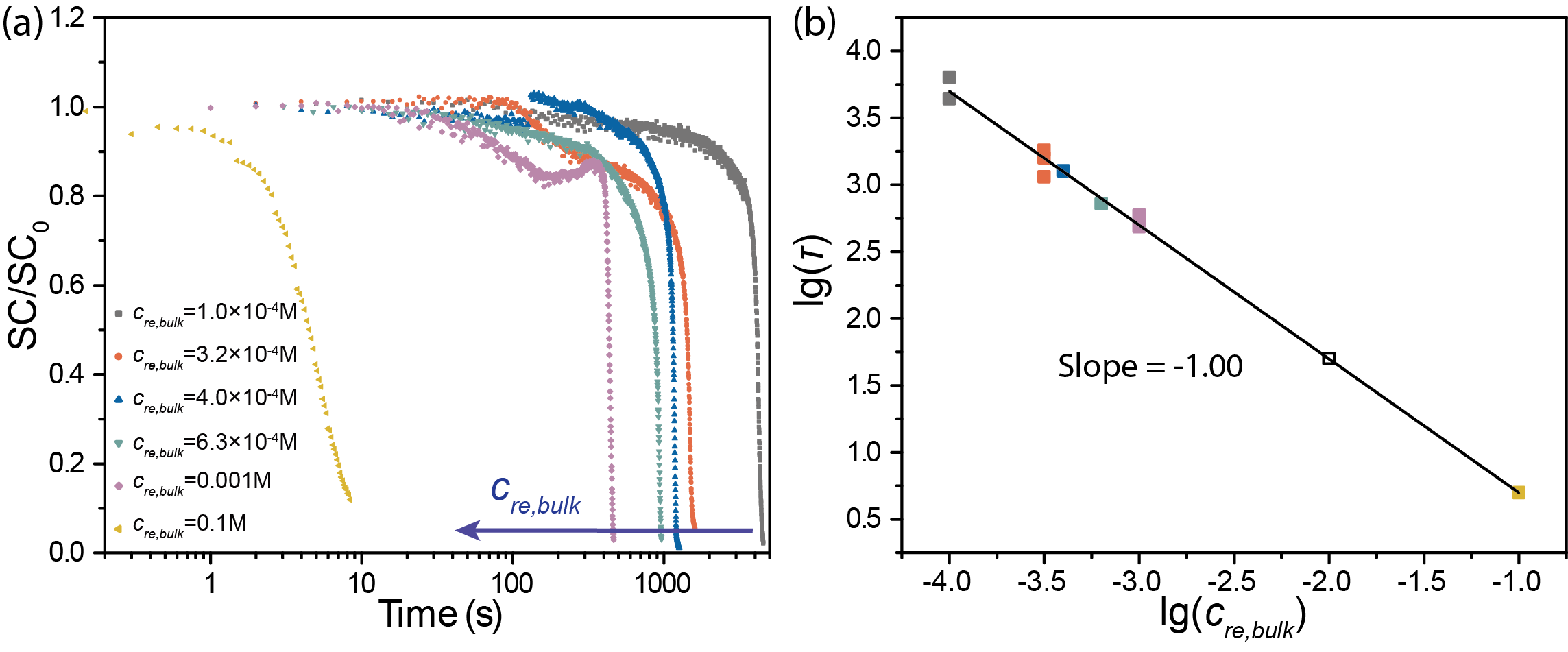}
  \caption{(a) Normalized surface coverage as function of time for droplets reacting with the alkaline solution of different concentrations. $SC$: surface coverage; $SC_0$: initial surface coverage. Data in (a) was obtained from the analysis of the images in (a)$-$(g) in Fig. \ref{fgr:pH1} (b) Droplet lifetime $\tau$ as function of alkaline concentration. $\tau$ on x-axis is the time required for $SC/SC_0$ to reach 0.1. Multiple data points for one concentration are the repeating experiments with initial surface coverage in a small range of variation ($3\%\sim 6\%$).}
  \label{fgr:pH2}
\end{figure*}

We now want to quantitatively obtain the dependence of the droplet lifetime $\tau$ on the concentration of the alkaline solution. 
Therefore, similar to our above analysis for the effects from the flow rate, the normalized surface coverage SC/SC$_0$ as function of time is presented in Figure \ref{fgr:pH2}(a).
Again, there is always a two-stage decrease in droplet surface coverage: not much change at the beginning and then a sudden decrease after a certain transition. The first stage of the droplet reaction is the shortest as the reactant concentration of the flow is the highest.

Fig. \ref{fgr:pH2}(b) shows the effective scaling relationship between the droplet lifetime and reactant concentration in the flow quantitatively.
The lifetime of the droplets effectively scales as $\tau$ $\propto$ ${c_{rc,bulk}}^{-1}$. The exponent of $-1.0$ fits the dissolution rate of the droplet radius as function of the  concentration of alkaline over three orders of magnitude ($0.1$M$-10^{-4}$M).

\subsection{Scaling analysis of droplet reaction with the flow}

In this subsection, we focus on the theoretical analysis of the coupled effects from the flow and the reactant concentration on the reaction rate of surface nanodroplets. We consider that the overall process of the droplet dissolution consists of four sub-steps: (i) mass transport of the reagent (alkali) in the flow, (ii) chemical neutralization at the droplet surface, (iii) desorption of the product from the droplet surface, and (iv) the transport of the product in the flow. 
The reaction equation is given as

\begin{equation}
\begin{aligned}
 CH_{3}(CH_{2})_{7}CHCH(CH_{2})_{7}COOH + OH^{-}\\
 \rightleftharpoons CH_{3}(CH_{2})_{7}CHCH(CH_{2})_{7}COO^{-} + H_{2}O.
 \end{aligned}
\end{equation}

In the droplet, the concentration of the acid is 100\% (i.e. pure oleic acid). The alkali in the flow must reach the droplet surface to react. The amount of reactants supplied by the flow to the droplet surface per unit time is proportional to the $Pe$ number of the reacting flow and the concentration of the reactant in the bulk. Meanwhile, the depletion of alkali, which results from the reaction in the boundary layer adjacent to the droplet surface, can be immediately replenished by the  influx from the flow due to the abundance of alkali in the flow. 

For given time, as the neutralization is a fast reaction, we assume that the concentration ratio of the free acid and the product reaches the dynamic equilibrium immediately.
The product concentration at interface $c_{pr,sur}$ is governed by the kinetics of the neutralization between oleic acid and sodium hydroxide.

\begin{equation}
 c_{pr,sur} \sim K_rc_{re,sur}.
\end{equation}
$K_r$ is the equilibrium constant of the forward chemical reaction (2), which is determined by the Gibbs free energy of the chemical reaction with the droplets, independent of Peclet number and reactant concentration. In the range of reactant concentration for all our experiments, the droplet surface can be fully converted to the product when the reaction takes place in a fixed environment without the flow.\cite{langmuir1936composition} However, under the flow condition, the product concentration at the interface $c_{pr,sur}$ is dependent on Peclet number, as the product is removed from the surface constantly by the flow.

We consider a simple mode in the analysis of a droplet with the shape of a spherical cap, assuming the droplet dissolves in a constant contact angle, same as the situation of dissolving nanodroplets in a flow reported in the literature.\cite{zhang2015,bao2018}
The mass loss rate $\dot m_{pr}$ of the product from the droplet is given by the product concentration gradient $\partial_r c_{pr}\lvert_{R}$ at the interface,

\begin{equation}
 \dot m_{pr} \propto \rho R^2\dot R \propto DR^2\partial_r c_{pr}\lvert_{R},
\end{equation}
where $R$ and $\rho$ are respectively the droplet radius and the density.

The concentration gradient at the droplet surface $\partial_r c_{pr}\lvert_{R}$ can be estimated from the product concentration difference between the oleate concentration in the flow $c_{pr,\infty}$ and at the interface $c_{pr,sur}$ and the thickness $\lambda$ of the concentration boundary layer.
Assuming the product concentration in the flow is negligibly small ($c_{pr,\infty}\approx 0$), we then obtain

\begin{equation}
 \partial_r c_{pr}\lvert_{r=R} \sim \frac{c_{pr,sur}-c_{pr,\infty}}{\lambda} \sim \frac{c_{pr,sur}}{\lambda}.
\end{equation}

Here the diffusive boundary layer of the product is assumed to be of Prandtl-Blasius-Pohlhausen-type for laminar flow, namely

\begin{equation}
 \lambda \sim \frac{R}{\sqrt{Pe}}.
\end{equation}

From equations (4-6), we obtain

\begin{equation}
\rho R\dot R \sim D\sqrt{Pe}c_{pr,sur}.
\end{equation}
From here we immediately deduce the scaling of the life time $\tau$ of a droplet of radius $R_0$, namely

\begin{equation}
    \tau\propto \frac{R_0^2}{D}\frac{\rho_o}{c_{pr,sur}}Pe^{-1/2}.
\end{equation}
Now the concentration of the product on the surface linearly depends on the supply rate of reactant by the flow, i.e. on $Pe$.

\begin{equation}
    c_{pr,sur}\propto PeK_rc_{re,bulk}.
\end{equation}
Coupling equation (8) and (9) leads the scaling law for the lifetime of droplets as function of the $Pe$ number and the concentration of the alkaline flow $C_{re,bulk}$ as
\begin{equation}
    \tau\propto Pe^{-3/2}(K_rc_{re,bulk})^{-1}.
\end{equation}

\subsection{Postponed reactions of downstream droplets}
\begin{figure*}
 \centering
 \includegraphics[height=20.5cm]{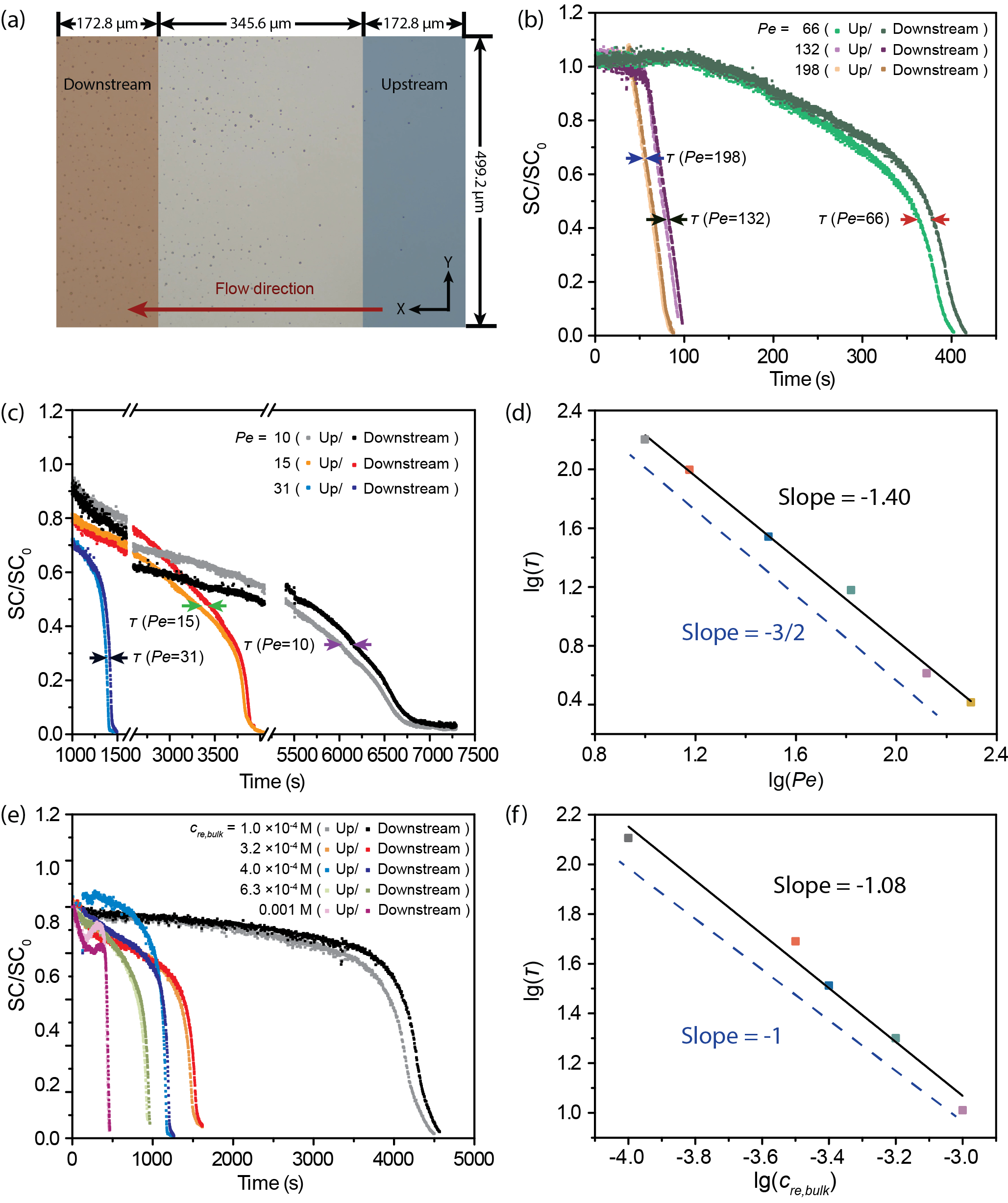}
 \caption{(a) An optical image showing the delayed downstream reaction, taken from the same experiment as Fig. \ref{fgr:Pe1}(d), at $t=1240 s$ (Pe=31, $c_{re,bulk}=4.0\times10^{-4}$M). The red arrow indicates the direction of the alkaline flow in X direction. The distance between the defined upstream region (blue shaded) and the downstream region (orange shaded) is 345.6 $\mu$m. Both regions span 172.8 $\mu$m along X direction. The width of the field of view is 499.2 $\mu$m. (b) and (c): the surface coverage normalized by its initial value ($SC/SC_0$) as function of time for different Peclet numbers in upstream (darker plots) and downstream (brighter plots) regions. The arrows indicate the measured delayed time $\tau$ for each flow rate between upstream and downstream. $\tau$ is the average time difference between upstream and downstream for the normalized surface coverage $SC/SC_0$ to reach 0.2, 0.3, 0.4, 0.5 and 0.6. (d) $\tau$ as function of Peclet number. 
 The black line was obtained by fitting the experimental data, while the blue dashed line was derived from the scaling analysis. (e) Normalized surface coverage ($SC/SC_0$) in upstream (darker symbols) and downstream (brighter symbols) regions as function of time at different alkaline concentrations. (f) $\tau$ as function of the alkaline concentration. The black line was obtained by fitting the experimental data, while the blue dashed line is from the scaling analysis.}
 \label{fgr:Delay}
\end{figure*}

Oleates produced by chemical reaction upstream can be transported along the flow to downstream droplets.
However, as the product is surface active, their adsorption on the surface of pristine droplets may influence the reaction of droplets along the path of the flow that is doped with the product. Here we examine the effect of the product from the upstream droplet reaction on the reaction with downstream droplets. 
While we reported the dissolution rate by comparing the change of the overall surface coverage in time, it is worth to note that the dissolution of droplets propagates along the flow direction. 
Figure \ref{fgr:Delay}(a) shows a snapshot taken from the video ($Pe=31$, $c_{re,bulk}=4\times 10^{-4}$M).
The whole snapshot was divided into several regions.
Two of these regions were defined as 'upstream region' and 'downstream region', respectively.
Although the droplets in the upstream (blue) region have fully disappeared, those in the downstream (brown) region still remain intact. 

It is remarkable that the product from the upstream region can completely inhibit the reaction of droplets in the downstream region. A possible explanation for the inhibition effect is that the product from the droplet reaction that happened upstream is transported to the surface of the droplets located downstream.
Owning to the high surface activity, the product in the flow readily adsorbs onto the droplet surface downstream, and inhibit the reaction and stabilize the coated droplets. 
After the upstream droplets were depleted, there is no more continuous supply of the product from the upstream region.
The product attached to the downstream droplets then is gradually flushed away with the fresh flow. The acid in the droplet becomes exposed to the alkali in the flow, initiated the shrinkage of the droplet from the reactions downstream. In this way, the droplets on the substrate progressively react with the alkali in the flow and dissolve gradually along the flow direction. 

Since the product concentration influences the the surface coverage of the reactant on the droplet surface and consequently the reaction rate, we are not able to obtain the kinetic constant based on the initial concentration of the reactant with product concentration varying with time. For the same reason, we cannot calculate the Damkohler number (Da), a dimensionless number relating the chemical reaction timescale to the mass transport. However, the strong influence of the flow rate on the lifetime of the droplets suggests that Da is larger than 1 under all of our flow rates. That is, the rate-limiting step is not the reaction rate but mass transport.

\par Fig. \ref{fgr:Delay}(b) and (c) compares how the normalized surface coverage (normalized by initial surface coverage) evolves upstream and downstream under different flow rates (the life time $\tau$ starts when the alkaline flow reaches the observed region). 
The gap between the upstream and downstream dissolution curves represents the timescale of the delay. 
The delay is shorter for the higher flow rate.
Fig. \ref{fgr:Delay}(d) shows the scaling relation between the timescale and the Peclet number. 
The result demonstrates that the timescale of the delay also obeys the scaling law $\tau$ $\propto$ $Pe^{-3/2}$.
The delay $\tau$ represents the time consumed for depleting all droplets in between the upstream and the downstream region.
The reason why the scaling for the delay again is $-3/2$ is simple. 
The downstream reactions speed up only when upstream droplets have already dissolved.
So that the delay between two different regions should be similar to the timescale for droplets between these two regions to react and dissolve.

The delay can also be quantified in the solution of different concentrations while the flow conditions of the solution flow are constant. Fig. \ref{fgr:Delay}(e) shows the temporal evolution of normalized surface coverage upstream and downstream for different NaOH concentrations.
The delay is shorter for the higher concentration of alkali. The relationship between the timescale for the delay and NaOH concentration seen in \ref{fgr:Delay}(f) also obeys the scaling law $\tau$ $\propto$ $c_{re,bulk}^{-1}$. 
This scaling is entirely consistent with the correlation between the lifetime of droplets and the alkali concentration in the flow shown in Fig. \ref{fgr:Pe2}.


The dissolution of non-reacting droplet array exposed to a flow of immiscible liquid has been studied in previous work.  It was found that not only the flow rate but also the location of the droplet in the array influence the rate of droplet dissolution.\cite{bao2018} Droplets at the corners and the edges of the arrays dissolve faster than the droplets surrounded by many neighbours due to collective effect in droplet dissolution. The collective effects are more pronounced at low flow rates, \cite{zhang2012mass,mustafa2016enhanced} resembling the delay of downstream droplet reactions observed in our present work. The lifetime of the non-reacting droplets followed the scaling law $\tau\propto Pe^{-1/2}$, clearly different from the case of reacting droplets in our cases. As shown here, the difference in the scaling relation with $Pe$
can be attributed to the effect of the transport of reactants with the flow.

\section{Conclusions}
We investigate the kinetics of chemical reaction between surface nanodroplets and the solute in a flow, and establish the relationships between the reaction rate of the droplets as revealed by their shrinkage and the flow rate and the reactant concentration in the bulk.
The reaction of the droplets becomes faster at higher flow rate or with a higher concentration of the reactant in the flow. The droplet reaction time scales with $\sim Pe^{-3/2}c_{re,bulk}^{-1}$. Enhanced transport of the reactant and of the product from the droplet surface by the flow contribute to accelerated kinetics of droplet reaction. Along the direction of the flow, the product from the upstream reaction postpones the reaction of the downstream droplets. 

As demonstrated in this work, even a simple acid-base reaction that takes place at the interface between the droplets and the immiscible flow can involve complicated mechanisms influenced by chemical kinetics, interface phenomena, convective and diffusive transport. The understanding presented in this work provides a useful insight into the design and control of droplet reactions in a broad range of applications, such as droplet-based sensing, heterogeneous catalysis or polymer particle synthesis.

\section*{Conflicts of interest}
There are no conflicts to declare.

\section*{Acknowledgements}
The project is supported by the Natural Science and Engineering Research Council of Canada (NSERC) and Future Energy Systems (Canada First Research
Excellence Fund). This research was undertaken, in part, thanks to funding from the Canada Research Chairs program.
This project is partly supported by Institute of Global Innovation Research in TUAT.
This work is partially funded by DL's ERC Advanced Grant DDD (No. 740479).


\balance


\bibliography{rsc} 

\providecommand*{\mcitethebibliography}{\thebibliography}
\csname @ifundefined\endcsname{endmcitethebibliography}
{\let\endmcitethebibliography\endthebibliography}{}
\begin{mcitethebibliography}{35}
\providecommand*{\natexlab}[1]{#1}
\providecommand*{\mciteSetBstSublistMode}[1]{}
\providecommand*{\mciteSetBstMaxWidthForm}[2]{}
\providecommand*{\mciteBstWouldAddEndPuncttrue}
  {\def\EndOfBibitem{\unskip.}}
\providecommand*{\mciteBstWouldAddEndPunctfalse}
  {\let\EndOfBibitem\relax}
\providecommand*{\mciteSetBstMidEndSepPunct}[3]{}
\providecommand*{\mciteSetBstSublistLabelBeginEnd}[3]{}
\providecommand*{\EndOfBibitem}{}
\mciteSetBstSublistMode{f}
\mciteSetBstMaxWidthForm{subitem}
{(\emph{\alph{mcitesubitemcount}})}
\mciteSetBstSublistLabelBeginEnd{\mcitemaxwidthsubitemform\space}
{\relax}{\relax}

\bibitem[Lohse and Zhang(2015)]{lohse2015surface}
D.~Lohse and X.~Zhang, \emph{Reviews of Modern Physics}, 2015, \textbf{87},
  981\relax
\mciteBstWouldAddEndPuncttrue
\mciteSetBstMidEndSepPunct{\mcitedefaultmidpunct}
{\mcitedefaultendpunct}{\mcitedefaultseppunct}\relax
\EndOfBibitem
\bibitem[Thiam \emph{et~al.}(2013)Thiam, Farese~Jr, and
  Walther]{thiam2013biophysics}
A.~R. Thiam, R.~V. Farese~Jr and T.~C. Walther, \emph{Nature Reviews Molecular
  Cell Biology}, 2013, \textbf{14}, 775--786\relax
\mciteBstWouldAddEndPuncttrue
\mciteSetBstMidEndSepPunct{\mcitedefaultmidpunct}
{\mcitedefaultendpunct}{\mcitedefaultseppunct}\relax
\EndOfBibitem
\bibitem[Li \emph{et~al.}(2017)Li, Mao, Leng, Ye, Sun, and
  Xu]{li2017enhancement}
X.~Li, Y.~Mao, K.~Leng, G.~Ye, Y.~Sun and W.~Xu, \emph{Microporous and
  Mesoporous Materials}, 2017, \textbf{254}, 114--120\relax
\mciteBstWouldAddEndPuncttrue
\mciteSetBstMidEndSepPunct{\mcitedefaultmidpunct}
{\mcitedefaultendpunct}{\mcitedefaultseppunct}\relax
\EndOfBibitem
\bibitem[Liu \emph{et~al.}(2002)Liu, Lee, Han, Chen, and Gan]{liu2002synthesis}
Z.~Liu, J.~Y. Lee, M.~Han, W.~Chen and L.~M. Gan, \emph{Journal of Materials
  Chemistry}, 2002, \textbf{12}, 2453--2458\relax
\mciteBstWouldAddEndPuncttrue
\mciteSetBstMidEndSepPunct{\mcitedefaultmidpunct}
{\mcitedefaultendpunct}{\mcitedefaultseppunct}\relax
\EndOfBibitem
\bibitem[Lin \emph{et~al.}(2016)Lin, Zhang, Li, and Deng]{lin2016magnetic}
Z.~Lin, Z.~Zhang, Y.~Li and Y.~Deng, \emph{Chemical Engineering Journal}, 2016,
  \textbf{288}, 305--311\relax
\mciteBstWouldAddEndPuncttrue
\mciteSetBstMidEndSepPunct{\mcitedefaultmidpunct}
{\mcitedefaultendpunct}{\mcitedefaultseppunct}\relax
\EndOfBibitem
\bibitem[Okubo \emph{et~al.}(1980)Okubo, Yamada, and
  Matsumoto]{okubo1980estimation}
M.~Okubo, A.~Yamada and T.~Matsumoto, \emph{Journal of Polymer Science: Polymer
  Chemistry Edition}, 1980, \textbf{18}, 3219--3228\relax
\mciteBstWouldAddEndPuncttrue
\mciteSetBstMidEndSepPunct{\mcitedefaultmidpunct}
{\mcitedefaultendpunct}{\mcitedefaultseppunct}\relax
\EndOfBibitem
\bibitem[Wang \emph{et~al.}(2019)Wang, Lin, Zhou, Xie, Song, Li, Huang, Huang,
  and Mann]{wang2019autonomic}
L.~Wang, Y.~Lin, Y.~Zhou, H.~Xie, J.~Song, M.~Li, Y.~Huang, X.~Huang and
  S.~Mann, \emph{Angewandte Chemie}, 2019, \textbf{131}, 1079--1083\relax
\mciteBstWouldAddEndPuncttrue
\mciteSetBstMidEndSepPunct{\mcitedefaultmidpunct}
{\mcitedefaultendpunct}{\mcitedefaultseppunct}\relax
\EndOfBibitem
\bibitem[Qian \emph{et~al.}(2019)Qian, Arends, and Zhang]{qian2019surface}
J.~Qian, G.~F. Arends and X.~Zhang, \emph{Langmuir}, 2019, \textbf{35},
  12583--12596\relax
\mciteBstWouldAddEndPuncttrue
\mciteSetBstMidEndSepPunct{\mcitedefaultmidpunct}
{\mcitedefaultendpunct}{\mcitedefaultseppunct}\relax
\EndOfBibitem
\bibitem[Bain \emph{et~al.}(2017)Bain, Sathyamoorthi, and
  Zare]{bain2017droplet}
R.~M. Bain, S.~Sathyamoorthi and R.~N. Zare, \emph{Angewandte Chemie}, 2017,
  \textbf{129}, 15279--15283\relax
\mciteBstWouldAddEndPuncttrue
\mciteSetBstMidEndSepPunct{\mcitedefaultmidpunct}
{\mcitedefaultendpunct}{\mcitedefaultseppunct}\relax
\EndOfBibitem
\bibitem[Kelly \emph{et~al.}(2007)Kelly, Baret, Taly, and
  Griffiths]{kelly2007miniaturizing}
B.~T. Kelly, J.-C. Baret, V.~Taly and A.~D. Griffiths, \emph{Chemical
  Communications}, 2007,  1773--1788\relax
\mciteBstWouldAddEndPuncttrue
\mciteSetBstMidEndSepPunct{\mcitedefaultmidpunct}
{\mcitedefaultendpunct}{\mcitedefaultseppunct}\relax
\EndOfBibitem
\bibitem[Wang \emph{et~al.}(2009)Wang, Yang, and Li]{wang2009efficient}
W.~Wang, C.~Yang and C.~M. Li, \emph{Small}, 2009, \textbf{5}, 1149--1152\relax
\mciteBstWouldAddEndPuncttrue
\mciteSetBstMidEndSepPunct{\mcitedefaultmidpunct}
{\mcitedefaultendpunct}{\mcitedefaultseppunct}\relax
\EndOfBibitem
\bibitem[Feng \emph{et~al.}(2018)Feng, Ueda, and Levkin]{feng2018droplet}
W.~Feng, E.~Ueda and P.~A. Levkin, \emph{Advanced Materials}, 2018,
  \textbf{30}, 1706111\relax
\mciteBstWouldAddEndPuncttrue
\mciteSetBstMidEndSepPunct{\mcitedefaultmidpunct}
{\mcitedefaultendpunct}{\mcitedefaultseppunct}\relax
\EndOfBibitem
\bibitem[Guardingo \emph{et~al.}(2016)Guardingo, Busqu{\'e}, and
  Ruiz-Molina]{guardingo2016reactions}
M.~Guardingo, F.~Busqu{\'e} and D.~Ruiz-Molina, \emph{Chemical Communications},
  2016, \textbf{52}, 11617--11626\relax
\mciteBstWouldAddEndPuncttrue
\mciteSetBstMidEndSepPunct{\mcitedefaultmidpunct}
{\mcitedefaultendpunct}{\mcitedefaultseppunct}\relax
\EndOfBibitem
\bibitem[Yan \emph{et~al.}(2016)Yan, Bain, and Cooks]{yan2016organic}
X.~Yan, R.~M. Bain and R.~G. Cooks, \emph{Angewandte Chemie International
  Edition}, 2016, \textbf{55}, 12960--12972\relax
\mciteBstWouldAddEndPuncttrue
\mciteSetBstMidEndSepPunct{\mcitedefaultmidpunct}
{\mcitedefaultendpunct}{\mcitedefaultseppunct}\relax
\EndOfBibitem
\bibitem[Lee \emph{et~al.}(2015)Lee, Banerjee, Nam, and
  Zare]{lee2015acceleration}
J.~K. Lee, S.~Banerjee, H.~G. Nam and R.~N. Zare, \emph{Quarterly Reviews of
  Biophysics}, 2015, \textbf{48}, 437--444\relax
\mciteBstWouldAddEndPuncttrue
\mciteSetBstMidEndSepPunct{\mcitedefaultmidpunct}
{\mcitedefaultendpunct}{\mcitedefaultseppunct}\relax
\EndOfBibitem
\bibitem[Lee \emph{et~al.}(2015)Lee, Kim, Nam, and Zare]{lee2015microdroplet}
J.~K. Lee, S.~Kim, H.~G. Nam and R.~N. Zare, \emph{Proceedings of the National
  Academy of Sciences}, 2015, \textbf{112}, 3898--3903\relax
\mciteBstWouldAddEndPuncttrue
\mciteSetBstMidEndSepPunct{\mcitedefaultmidpunct}
{\mcitedefaultendpunct}{\mcitedefaultseppunct}\relax
\EndOfBibitem
\bibitem[Wei \emph{et~al.}(2017)Wei, Wleklinski, Ferreira, and
  Cooks]{wei2017reaction}
Z.~Wei, M.~Wleklinski, C.~Ferreira and R.~G. Cooks, \emph{Angewandte Chemie},
  2017, \textbf{129}, 9514--9518\relax
\mciteBstWouldAddEndPuncttrue
\mciteSetBstMidEndSepPunct{\mcitedefaultmidpunct}
{\mcitedefaultendpunct}{\mcitedefaultseppunct}\relax
\EndOfBibitem
\bibitem[Griffith and Vaida(2012)]{griffith2012situ}
E.~C. Griffith and V.~Vaida, \emph{Proceedings of the National Academy of
  Sciences}, 2012, \textbf{109}, 15697--15701\relax
\mciteBstWouldAddEndPuncttrue
\mciteSetBstMidEndSepPunct{\mcitedefaultmidpunct}
{\mcitedefaultendpunct}{\mcitedefaultseppunct}\relax
\EndOfBibitem
\bibitem[Wiebenga-Sanford \emph{et~al.}(2016)Wiebenga-Sanford, DiVerdi,
  Rithner, and Levinger]{wiebenga2016nanoconfinement}
B.~P. Wiebenga-Sanford, J.~DiVerdi, C.~D. Rithner and N.~E. Levinger, \emph{The
  Journal of Physical Chemistry Letters}, 2016, \textbf{7}, 4597--4601\relax
\mciteBstWouldAddEndPuncttrue
\mciteSetBstMidEndSepPunct{\mcitedefaultmidpunct}
{\mcitedefaultendpunct}{\mcitedefaultseppunct}\relax
\EndOfBibitem
\bibitem[Lee \emph{et~al.}(2018)Lee, Samanta, Nam, and
  Zare]{lee2018spontaneous}
J.~K. Lee, D.~Samanta, H.~G. Nam and R.~N. Zare, \emph{Nature Communications},
  2018, \textbf{9}, 1562\relax
\mciteBstWouldAddEndPuncttrue
\mciteSetBstMidEndSepPunct{\mcitedefaultmidpunct}
{\mcitedefaultendpunct}{\mcitedefaultseppunct}\relax
\EndOfBibitem
\bibitem[Nam \emph{et~al.}(2017)Nam, Lee, Nam, and Zare]{nam2017abiotic}
I.~Nam, J.~K. Lee, H.~G. Nam and R.~N. Zare, \emph{Proceedings of the National
  Academy of Sciences}, 2017, \textbf{114}, 12396--12400\relax
\mciteBstWouldAddEndPuncttrue
\mciteSetBstMidEndSepPunct{\mcitedefaultmidpunct}
{\mcitedefaultendpunct}{\mcitedefaultseppunct}\relax
\EndOfBibitem
\bibitem[Fallah-Araghi \emph{et~al.}(2014)Fallah-Araghi, Meguellati, Baret,
  El~Harrak, Mangeat, Karplus, Ladame, Marques, and
  Griffiths]{fallah2014enhanced}
A.~Fallah-Araghi, K.~Meguellati, J.-C. Baret, A.~El~Harrak, T.~Mangeat,
  M.~Karplus, S.~Ladame, C.~M. Marques and A.~D. Griffiths, \emph{Physical
  Review Letters}, 2014, \textbf{112}, 028301\relax
\mciteBstWouldAddEndPuncttrue
\mciteSetBstMidEndSepPunct{\mcitedefaultmidpunct}
{\mcitedefaultendpunct}{\mcitedefaultseppunct}\relax
\EndOfBibitem
\bibitem[Vaida(2017)]{vaida2017prebiotic}
V.~Vaida, \emph{Proceedings of the National Academy of Sciences}, 2017,
  \textbf{114}, 12359--12361\relax
\mciteBstWouldAddEndPuncttrue
\mciteSetBstMidEndSepPunct{\mcitedefaultmidpunct}
{\mcitedefaultendpunct}{\mcitedefaultseppunct}\relax
\EndOfBibitem
\bibitem[Urban(2014)]{urban2014compartmentalised}
P.~L. Urban, \emph{New Journal of Chemistry}, 2014, \textbf{38},
  5135--5141\relax
\mciteBstWouldAddEndPuncttrue
\mciteSetBstMidEndSepPunct{\mcitedefaultmidpunct}
{\mcitedefaultendpunct}{\mcitedefaultseppunct}\relax
\EndOfBibitem
\bibitem[Carroll and Hidrovo(2013)]{carroll2013experimental}
B.~Carroll and C.~Hidrovo, \emph{Heat Transfer Engineering}, 2013, \textbf{34},
  120--130\relax
\mciteBstWouldAddEndPuncttrue
\mciteSetBstMidEndSepPunct{\mcitedefaultmidpunct}
{\mcitedefaultendpunct}{\mcitedefaultseppunct}\relax
\EndOfBibitem
\bibitem[Nakatani \emph{et~al.}(1995)Nakatani, Suto, Wakabayashi, Kim, and
  Kitamura]{nakatani1995direct}
K.~Nakatani, T.~Suto, M.~Wakabayashi, H.-B. Kim and N.~Kitamura, \emph{The
  Journal of Physical Chemistry}, 1995, \textbf{99}, 4745--4749\relax
\mciteBstWouldAddEndPuncttrue
\mciteSetBstMidEndSepPunct{\mcitedefaultmidpunct}
{\mcitedefaultendpunct}{\mcitedefaultseppunct}\relax
\EndOfBibitem
\bibitem[Nakatani \emph{et~al.}(1995)Nakatani, Chikama, Kim, and
  Kitamura]{nakatani1995droplet}
K.~Nakatani, K.~Chikama, H.-B. Kim and N.~Kitamura, \emph{Chemical Physics
  Letters}, 1995, \textbf{237}, 133--136\relax
\mciteBstWouldAddEndPuncttrue
\mciteSetBstMidEndSepPunct{\mcitedefaultmidpunct}
{\mcitedefaultendpunct}{\mcitedefaultseppunct}\relax
\EndOfBibitem
\bibitem[Nakatani \emph{et~al.}(1996)Nakatani, Wakabayashi, Chikama, and
  Kitamura]{nakatani1996electrochemical}
K.~Nakatani, M.~Wakabayashi, K.~Chikama and N.~Kitamura, \emph{The Journal of
  Physical Chemistry}, 1996, \textbf{100}, 6749--6754\relax
\mciteBstWouldAddEndPuncttrue
\mciteSetBstMidEndSepPunct{\mcitedefaultmidpunct}
{\mcitedefaultendpunct}{\mcitedefaultseppunct}\relax
\EndOfBibitem
\bibitem[Lu \emph{et~al.}(2016)Lu, Peng, and Zhang]{lu2016influence}
Z.~Lu, S.~Peng and X.~Zhang, \emph{Langmuir}, 2016, \textbf{32},
  1700--1706\relax
\mciteBstWouldAddEndPuncttrue
\mciteSetBstMidEndSepPunct{\mcitedefaultmidpunct}
{\mcitedefaultendpunct}{\mcitedefaultseppunct}\relax
\EndOfBibitem
\bibitem[Peng \emph{et~al.}(2015)Peng, Lohse, and Zhang]{peng2015spontaneous}
S.~Peng, D.~Lohse and X.~Zhang, \emph{ACS Nano}, 2015, \textbf{9},
  11916--11923\relax
\mciteBstWouldAddEndPuncttrue
\mciteSetBstMidEndSepPunct{\mcitedefaultmidpunct}
{\mcitedefaultendpunct}{\mcitedefaultseppunct}\relax
\EndOfBibitem
\bibitem[Bao \emph{et~al.}({2018})Bao, Spandan, Yang, Dyett, Verzicco, Lohse,
  and Zhang]{bao2018}
L.~Bao, V.~Spandan, Y.~Yang, B.~Dyett, R.~Verzicco, D.~Lohse and X.~Zhang,
  \emph{{Lab on a Chip}}, {2018}, \textbf{{18}}, 1066--1074\relax
\mciteBstWouldAddEndPuncttrue
\mciteSetBstMidEndSepPunct{\mcitedefaultmidpunct}
{\mcitedefaultendpunct}{\mcitedefaultseppunct}\relax
\EndOfBibitem
\bibitem[Lessel \emph{et~al.}(2015)Lessel, B{\"a}umchen, Klos, H{\"a}hl,
  Fetzer, Paulus, Seemann, and Jacobs]{lessel2015}
M.~Lessel, O.~B{\"a}umchen, M.~Klos, H.~H{\"a}hl, R.~Fetzer, M.~Paulus,
  R.~Seemann and K.~Jacobs, \emph{Surface and Interface Analysis}, 2015,
  \textbf{47}, 557--564\relax
\mciteBstWouldAddEndPuncttrue
\mciteSetBstMidEndSepPunct{\mcitedefaultmidpunct}
{\mcitedefaultendpunct}{\mcitedefaultseppunct}\relax
\EndOfBibitem
\bibitem[Langmuir and Schaefer(1936)]{langmuir1936composition}
I.~Langmuir and V.~J. Schaefer, \emph{Journal of the American Chemical
  Society}, 1936, \textbf{58}, 284--287\relax
\mciteBstWouldAddEndPuncttrue
\mciteSetBstMidEndSepPunct{\mcitedefaultmidpunct}
{\mcitedefaultendpunct}{\mcitedefaultseppunct}\relax
\EndOfBibitem
\bibitem[Zhang \emph{et~al.}(2012)Zhang, Yang, and Mao]{zhang2012mass}
J.~Zhang, C.~Yang and Z.-S. Mao, \emph{AIChE Journal}, 2012, \textbf{58},
  3214--3223\relax
\mciteBstWouldAddEndPuncttrue
\mciteSetBstMidEndSepPunct{\mcitedefaultmidpunct}
{\mcitedefaultendpunct}{\mcitedefaultseppunct}\relax
\EndOfBibitem
\bibitem[Mustafa \emph{et~al.}(2016)Mustafa, Erten, Ayaz, Kay{\i}ll{\i}oğlu,
  Eser, Eryürek, Irfan, Muradoglu, Tanyeri, and Kiraz]{mustafa2016enhanced}
A.~Mustafa, A.~Erten, R.~M.~A. Ayaz, O.~Kay{\i}ll{\i}oğlu, A.~Eser,
  M.~Eryürek, M.~Irfan, M.~Muradoglu, M.~Tanyeri and A.~Kiraz,
  \emph{Langmuir}, 2016, \textbf{32}, 9460--9467\relax
\mciteBstWouldAddEndPuncttrue
\mciteSetBstMidEndSepPunct{\mcitedefaultmidpunct}
{\mcitedefaultendpunct}{\mcitedefaultseppunct}\relax
\EndOfBibitem
\end{mcitethebibliography}
\bibliographystyle{rsc} 

\end{document}